\documentclass[runningheads]{llncs}
\usepackage{graphicx}
\usepackage{multirow}
\usepackage{amssymb}
\usepackage[misc]{ifsym}
\begin{document}
\title{Domain Composition and Attention for Unseen-Domain Generalizable Medical Image Segmentation}
\titlerunning{Domain Composition and Attention for Unseen-Domain Segmentation}
%
\author{Ran Gu\inst{1, 3, }\thanks{Ran Gu and Jingyang Zhang contributed equally. The work was done during their internship at SenseTime.} \and
Jingyang Zhang\inst{2, 3, \star} \and
Rui Huang\inst{3} \and 
Wenhui Lei\inst{1} \and
Guotai Wang\inst{1, }$^{\textrm{\Letter}}$ \and
Shaoting Zhang\inst{1, 3}
}

%
\authorrunning{R. Gu et al.}
%
\institute{School of Mechanical and Electrical Engineering, University of Electronic Science and Technology of China, Chengdu, China
\and
School of Biomedical Engineering, Shanghai Jiao Tong University, Shanghai, China \and 
SenseTime Research, Shanghai, China \\
\email{guotai.wang@uestc.edu.cn}}
%
\maketitle              
\begin{abstract}
Domain generalizable model is attracting increasing attention in medical image analysis since data is commonly acquired from different institutes with various imaging protocols and scanners. To tackle this challenging domain generalization problem, we propose a Domain Composition and Attention-based network (DCA-Net) to improve the ability of domain representation and generalization. 
First, we present a domain composition method that represents one certain domain by a linear combination of a set of basis representations (i.e., a representation bank). Second, a novel plug-and-play parallel domain preceptor is proposed to learn these basis representations and we introduce a divergence constraint function to encourage the basis representations are as divergent as possible. Then, a domain attention module is proposed to learn the linear combination coefficients of the basis representations. The result of liner combination is used to calibrate the feature maps of an input image, which enables the model to generalize to different and even unseen domains. We validate our method on public prostate MRI dataset acquired from six different institutions with apparent domain shift. Experimental results show that our proposed model can generalizes well on different and even unseen domains and it outperforms state-of-the-art methods on the multi-domain prostate segmentation task. Code is available at \url{https://github.com/HiLab-git/DCA-Net}.

\end{abstract}
\section{Introduction}
Deep learning with Convolutional Neural Networks (CNNs) have achieved remarkable performance for medical image segmentation\cite{litjens2014evaluation,ronneberger2015u,shen2017deep}. Their achievements heavily rely on the training process with images from a similar domain distribution for test images. However, for clinical model deployment, the training and test images are commonly acquired from different sites (hospitals) with different imaging protocols and scanner vendors, leading to evident data distribution discrepancy between them and thus substantial performance degradation at test time~\cite{liu2020ms,ganin2016domain,kamnitsas2017unsupervised}. Recently, Domain Adaptation (DA) methods~\cite{ganin2016domain,dou2018unsupervised,perone2019} provide a promising way to solve this problem by aligning the source and target domain distributions. While, most of the DA methods require repeated model re-training with involvement of images from the target domain, which is time-consuming and impractical when the images from a new domain are not available for training in clinical applications~\cite{chen2018semantic}. Therefore, it is desirable to enable a model to generalize well to unseen domains, where no images in the new domain are available for retraining the model. Nowadays, domain generalization is still a challenging task but full of practical clinical significance.


Recently, there are some researches have tried to solving the domain generalization problem~\cite{li2018learning,li2018domain,balaji2018metareg}, and some methods has been applied on medical imaging. From the aspect of data, Zhang et al.~\cite{zhang2020generalizing} proposed a data augmentation-based method assuming that the shift between source and target domains could be simulated by applying extensive data augmentation on a single source domain. 
However, the configuration of augmentation operations, such as the type and the number of transformation, requires empirical settings and even data-specific modifications. From the aspect of algorithm, Dou et al.~\cite{dou2019domain} proposed a meta-learning paradigm benefited from its model-agnostic manner and providing a flexible way to solve the domain generalization problem. It splits a set of source domains into meta-train and meta-test subsets, and adopts a gradient-based meta-optimization that iteratively updates model parameters to improve performance on both meta-train and meta-test datasets. Subsequently, Liu et al.~\cite{liu2020shape} further introduced shape constraint based on meta-learning for generalizable prostate MRI segmentation. Although meta-learning exhibits remarkable performance for domain generalization~\cite{li2018learning}, it still has two-big limitations. First, the meta-optimization process is highly time-consuming since all potential splitting results of meta-train and meta-test should be considered and involved during training. The process would be even more complicated and hard for convergence when dealing with a larger set of source domains. Second, as meta-learning is used to optimize a unidirectional transfer from meta-train to meta-test domains, it still needs to distinguish distribution discrepancy among source domains by domain-specific statistics, which brings more restricts for data acquirement and storage in clinical practice. And from the aspect of self-ensemble, Wang et al.~\cite{wang2020dofe} introduced a domain-oriented feature embedding framework that dynamically enriches the image features with additional domain prior knowledge learned from multi-source domains to make the semantic features more discriminative. However, ~\cite{wang2020dofe} extracts one domain prior knowledge from each of the $K$ source domains, where they are limited by one-to-one correspondence. Meanwhile, it simply takes an average of features from one domain as its prior knowledge, which may lead to redundancy of prior vectors when two domains are similar and limit the representation power.

Nowadays, attention mechanism is increasingly used because of its powerful self-expression ability, and plays a great role in semantic segmentation~\cite{chen2016attention,oktay2018attention,gu2021comprehensive}. Inspired by channel attention~\cite{hu2018se}, Wang et al.~\cite{wang2019towards} proposed a domain attention method to construct a universal model for multi-site object detection, leading to a striking performance. However, this method was only validated with domains presented for training, and its generalizablity to unseen domains was not explored.
For multi-site images, the segmentation model is expected to achieve high performance not only on diffident domains that have been seen during training, but also on unseen domains that are often involved in the model deployment stage. To deal with this problem, we propose a Domain Composition and Attention-based Network (DCA-Net) for generalizable multi-site medical image segmentation. 

The contributions of our work are: 1) we propose a novel Domain Composition and Attention (DCA) method to represent a certain (seen or unseen) domain by a linear combination of a set of basis domain representations, so that the generalizablility of segmentation models are improved. 2) We design a Parallel Domain Preceptor (PDP) to learn the basis domain representations, and propose a divergence constraint function to encourage the basis domain representations to be divergent, which improves the domain composition power for generalization. 3) We propose a network using DCA (i.e., DCA-Net) for generalizable segmentation, and experimental results on multi-site prostate MRI segmentation showed that our DCA-Net outperformed state-of-the-art domain generalization methods in the same experimental settings.
\section{Methods}
\begin{figure}
    \centering
    \includegraphics[width=\textwidth]{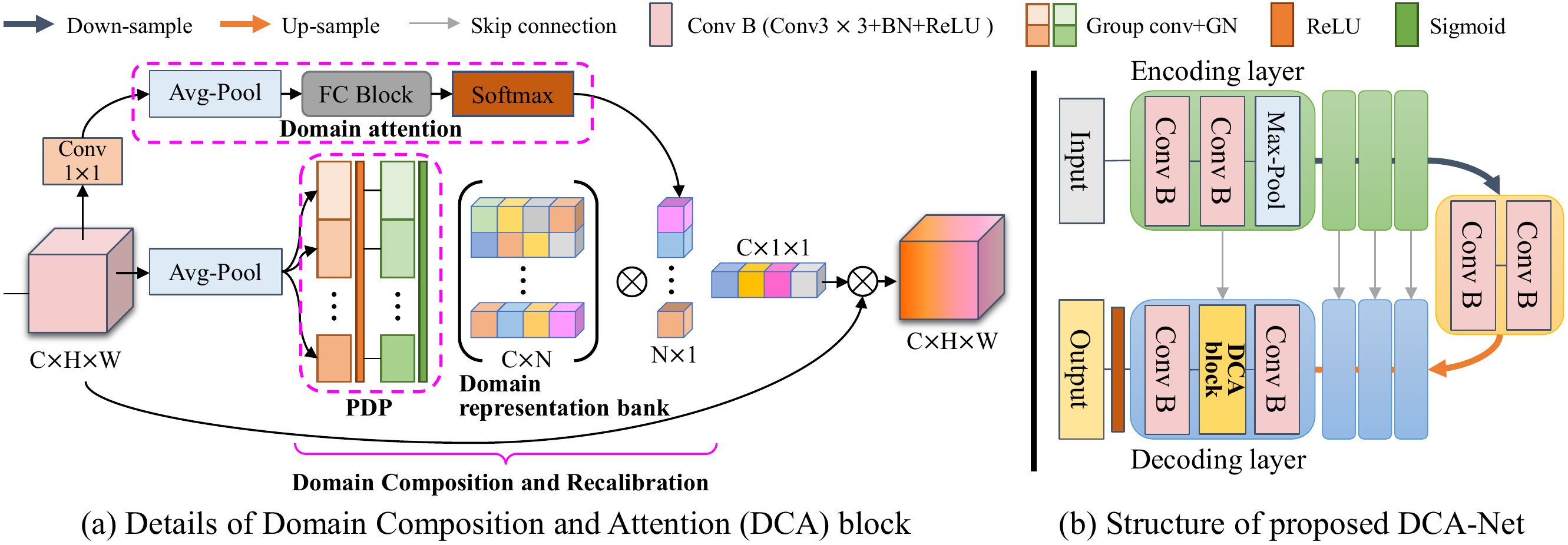}
    \caption{Overview of the proposed Domain Composition and Attention (DCA) method for multi-site domain generalizable segmentation. (a) Details of the proposed DCA block with a Parallel Domain Preceptor (PDP) and a domain attention module for domain composition. (b) Structure of DCA-Net, which combines the DCA block with the decoder of a U-Net backbone.}
    \label{fig1}
\end{figure}

Let $\mathcal{D} = \{ \mathcal{D}_1, \mathcal{D}_2, ..., \mathcal{D}_M \}$ be the set of $M$ source domains. The $k$-th image and its label in domain $D_m$ are denoted as $\mathit{x^{k}_m}$ and $\mathit{y^{k}_m}$, respectively. An unseen target domain for testing is denoted as $\mathcal{D}_t$. To achieve robust segmentation results for both source domains and unseen target domains, we proposed a Domain Composition and Attention (DCA) method to recalibrate the feature maps of an input image to obtain high segmentation performance across different domains. Inspired by shape composition~\cite{zhang2012sparse}, our domain composition represents a certain domain by a linear combination of elements in a domain representation bank consisting of a set of basis representations. As shown in Fig.~\ref{fig1}(a), the domain representation bank is learned by a Parallel Domain Preceptor (PDP) and the linear combination coefficients are learned by a domain attention module. The DCA block is combined with a U-Net backbone to achieve generalizability across different domains, which is named as DCA-Net and shown in Fig.~\ref{fig1} (b).  In the following, we frist introduce our DCA block and then describe the DCA-Net.

\subsection{Domain Composition and Attention Block} As shown in Fig.~\ref{fig1}(a), the goal of our DCA block is to calibrate a feature map $F$ of an image from a certain domain adaptively to neutralize its domain bias for improving the generalizability. DCA includes a PDP module for constructing a domain representation bank and a domain attention module for domain composition, the output of which is used for the feature neutralization. 

\subsubsection{Parallel Domain Preceptor (PDP).} The PDP is used to learn a set of $N$ basis representations in the domain representation bank. We use one basis preceptor for each basis representation, which is designed to capture a specific domain information that are different from the others. For the input feature map $F$ with $C$ channels, its avg-pooled result is denoted as $f\in \mathbb{R}^{C\times 1}$.  The $n$-th basis preceptor $p_n$ converts $f$ into the $n$-th basis representation $p_n(f)\in \mathbb{R}^{C\times 1}$, thus the domain representation bank can be donated as $\mathcal{B} = \{p_{1}(f), p_{2}(f), ..., p_{N}(f)\}$. As the basis preceptors are independent, they can be implemented by a set of parallel subnetworks. To improve the efficiency, we propose to use a Grouped Convolution (GC) subnetwork as the structure for PDP, where each group corresponds to a basis preceptor. Specifically, the GC subnetwork has two convolutional layers each with $N$ groups, and they are followed by group normalization + ReLU and group normalization + sigmoid, respectively, as illustrated in Fig.~\ref{fig1}(a).  Thus, the output of group $n$ obtains $p_n(f)$. As there is no interactions between different groups in GC, the $N$ preceptors (groups) are independent to capture different basis representations.

\subsubsection{Divergence Constraint for Representation Bank.}
\label{L_drd}
To encourage the preceptors to capture different domain information, we propose a divergence constraint function $\mathcal{L}_{div}$ that enforces outputs of the $N$ preceptors to be different from each other, which makes DCA be capable of capturing a wide-range domain representations. During training, we randomly choose out three basis preceptors $p_i$, $p_j$, $p_k$ and minimize the average consistency between each pair of them. 
\begin{equation}
    \mathcal{L}_{div} = 1-\big[\frac{1}{3}\Big(||p_i(f)-p_j(f)||^2_2 + ||p_i(f)-p_k(f)||^2_2 + ||p_j(f)-p_k(f)||^2_2\Big)\big]^\frac{1}{2}
\end{equation}
where $||\cdot||_2$ is the $L2$ norm, and $p_i$, $p_j$, $p_k$ ($i\ne j \ne k$) are randomly selected in each iteration.
As a result, the basis representations are encouraged to be divergent and $\mathcal{B}$ will have a strong representation ability with its diverse elements. 

\subsubsection{Domain Attention Module.} With the basis domain representations, we represent a corresponding domain calibration vector $\alpha(f)\in \mathbb{R}^{C\times 1}$ for the input feature map $F$ as a linear combination of them:  
\begin{equation}\label{eq:compose}
\alpha(f) = \beta_1 p_1(f) + \beta_2 p_2(f) + ... + \beta_Np_N(f)
\end{equation}
where $\beta_n$ is the coefficient for $p_n(f)$. The coefficient vector $\beta$=($\beta_1$,  $\beta_2$, ...,  $\beta_N$) is adaptively predicted by a domain attention module, as shown in Fig.~\ref{fig1}(a). For the input feature map $F$, we use a $1\times 1$ convolution followed by avg-pooling in a second branch to convert it to a vector, which is then followed by a Fullly Connected (FC) block and a softmax, where the last layer of FC has a length of $N$, thus the output of the domain attention module is taken as the coefficient vector $\beta$. Given $\beta$, we obtain $\alpha(f)$ based on Eq.~\ref{eq:compose}, and the output of DAC is $\hat{F} = F \otimes \alpha(f)$, where $\otimes$ means tensor multiplication with broadcasting.



\subsection{DCA-Net}
Without loss of generality, we chose the powerful U-Net~\cite{ronneberger2015u} as the backbone for its simple structure and good performance for segmentation. We use a DCA in each convolutional block at different levels in the decoder of U-Net, as shown in Fig.~\ref{fig1} (b). Thus, in our DCA-Net, each convolutional block in the decoder consists of two convolutional layers that are connected by the DCA. During training, we combine a standard segmentation loss (i.e., Dice loss) $L_{seg}$ with deep supervision to improve the performance. 
We also adopt the shape compactness constraint loss $\mathcal{L}_{comp}$ suggested by Liu et  al.~\cite{liu2020shape} as regularization to improve the robustness. Considering that we also need a divergence constraint function $\mathcal{L}_{div}$ as explained above, the overall loss function for our DCA-Net is: 
\begin{equation}
    \mathcal{L}_{all} = \mathcal{L}_{seg} + \lambda_{1}\mathcal{L}_{comp} + \lambda_{2}\mathcal{L}_{div}
\end{equation}
where $\lambda_{1}$ and $\lambda_{2}$ are the weights of $L_{comp}$ and $L_{div}$, respectively.

\section{Experiments and Results}
\subsubsection{Datasets and Implementation.} For experiments, we used the well-organized multi-site T2-weighted MRI dataset\footnote{https://liuquande.github.io/SAML/} for prostate segmentation~\cite{liu2020shape} which was acquired from 6 institutes with different patient numbers and acquisition protocols: Data of Site A and B are from NCI-ISBI13 dataset containing 30 samples each; data of Site C are from I2CVB dataset containing 19 samples; data of Site D, E and F are from PROMISE12 dataset containing 13, 12 and 12 samples, respectively. The images were preprocessed in the same way as SAML~\cite{liu2020shape}.  To assess the generalizability to unseen domains, we follow the leave-one-domain-out strategy in SAML~\cite{liu2020shape}, where each time one domain is used as unseen and the others are used for seen domains. All the images in the seen domains are used for training, and we split the unseen domain into  20\%  and 80\% at patient level  for validation and testing, respectively.

Our DCA-Net with UNet as backbone~\cite{ronneberger2015u} was implemented in 2D due to the large variance on through-plane spacing among different sites. The encoder is kept the original settings with the channel numbers of 16, 32, 64, 128 and 256 at five scales, respectively~\cite{ronneberger2015u}. We concatenate every three slices into an input with three channels, and use DCA-Net to predict the segmentation in the central slice. The loss function weights $\lambda_{1}$ and $\lambda_{2}$ are set as 1.0 and 0.1, respectively. The segmentation model was trained using Adam optimizer and the learning rate was $5e^{-4}$. We trained 20k iterations with batch size of 4 and the basis representation number $N$ was 8 in the domain representation bank. Training was implemented on one NVIDIA Geforce GTX 1080 Ti GPU.
For fair comparison, we kept the most parameters be same as those in SAML~\cite{liu2020shape}. We adopt Dice score (Dice) and Average Surface Distance (ASD) as the evaluation metrics.
\subsubsection{Generalizability of DCA-Net on Unseen Domains.} 
We used the `DeepAll' as a baseline which means training all source domains jointly using a standard supervised learning strategy with U-Net~\cite{ronneberger2015u} and testing on the unseen domain. Furthermore, we compared our DCA-Net with several state-of-the-art generalization methods, including the data-augmentation based method (BigAug)~\cite{li2018domain}, a meta-learning based domain generalization method (MASF)~\cite{dou2019domain}, and a shape-aware meta-learning method (SAML)~\cite{liu2020shape}.

\begin{figure}
    \centering
    \includegraphics[width=0.92\textwidth]{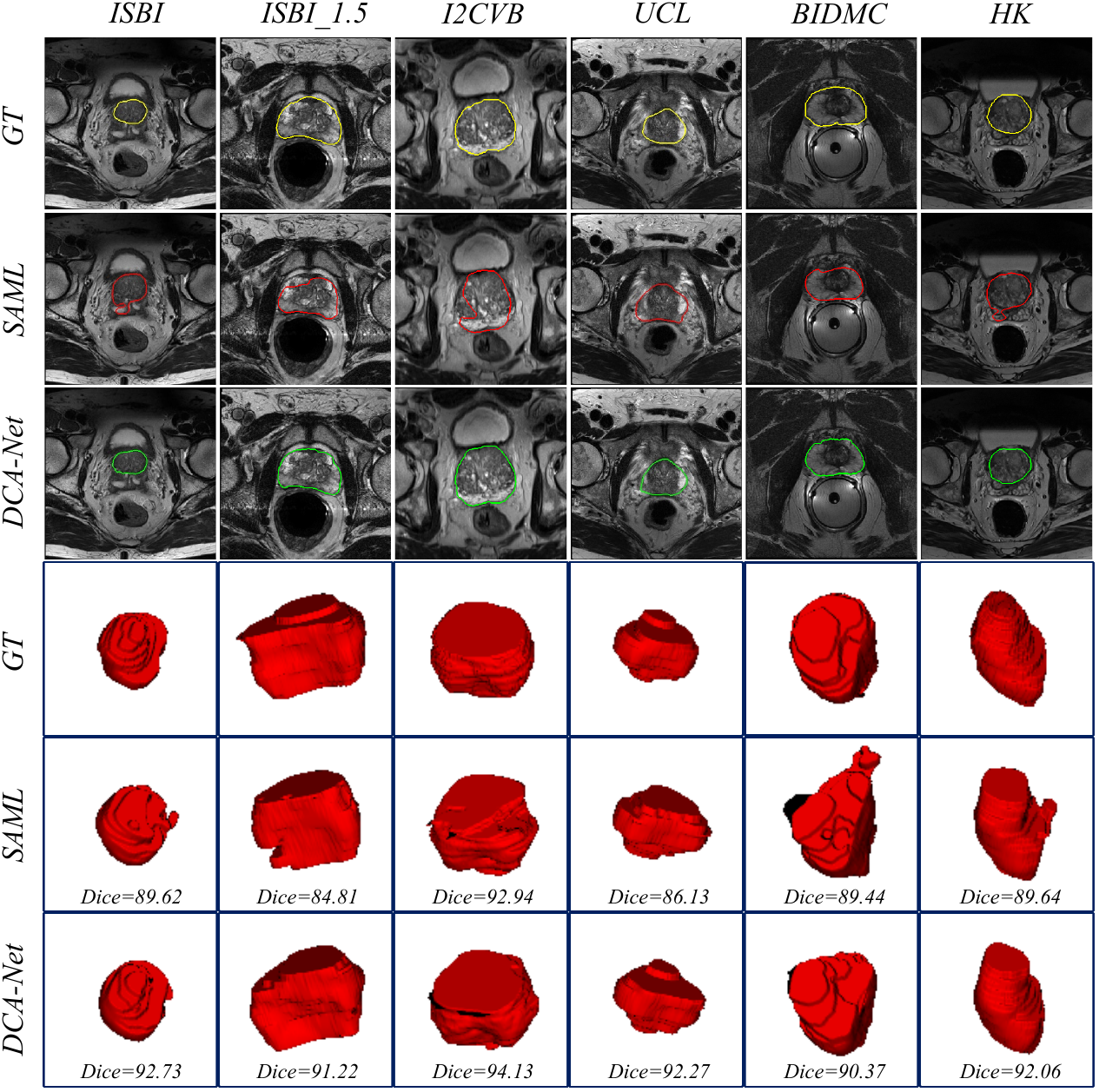}
    \caption{Visual comparison between DCA-Net and SAML~\cite{liu2020shape}. The first and last three rows are corresponding 2D and 3D visualizations, respectively. Each column shows the results of one domain used for testing and the others for training. }
    \label{fig4}
\end{figure}

The `Unseen' section of Table~\ref{tab2} shows the quantitative evaluation results, where the values of of BigAug, MASF and SAML are from \cite{liu2020shape} due to the same experimental setting as ours. It demonstrates that BigAug and MASF are effective for domain generalization, which shows advantage over the lower bound DeepAll. 
SAML using shape-aware meta-learning achieved the best performance among existing methods, which gets and average Dice of 87.16\% and average ASD of 1.58 $mm$ across the 6 domains. Remarkably, our proposed DCA-Net achieved higher performance in terms of average Dice score and ASD, which are 88.16\% and 1.29 $mm$, respectively. We achieved the highest performance in 5 out of the 6 domains. It is noted that though DCA-Net has a lower Dice score than SAML on the I2CVB site, it has a lower ASD, which means the contour of segmentation obtained by DCA-Net is closer to the ground truth. Fig.~\ref{fig4} provides 2D and 3D visual comparisons between SAML and DCA-Net for images from the 6 domains respectively. The 3D visualization shows that the results of our approach are closer to the ground truth. The 2D visualization shows that SAML has more over- and mis-segmented regions than ours.
\begin{table}
    \small
    \centering
    \caption{Generalizable segmentation performance of various methods on Dice (\%) and ASD ($mm$). $^*$ means training with $\mathcal{L}_{seg}$ and $\mathcal{L}_{comp}.$}\label{tab2}
    \scalebox{0.78}{\begin{tabular}{l|l|l r|l r|l r|l r|l r|l r|l r}
    \hline
    Scene & Method & \multicolumn{2}{c}{ISBI} & \multicolumn{2}{|c|}{ISBI1.5} & \multicolumn{2}{|c|}{I2CVB} & \multicolumn{2}{|c|}{UCL} & \multicolumn{2}{|c|}{BIDMC} & \multicolumn{2}{|c|}{HK} & \multicolumn{2}{|c}{Average} \\
    \hline
    \multirow{3}{0.4in}{Seen} & Intra-site & 89.27 & 1.41 & 88.17 & 1.35 & 88.29 & 1.56 & 83.23 & 3.21 & 83.67 & 2.93 & 85.43 & 1.91 & 86.34 & 2.06 \\
     & DeepAll$^*$ & 91.37 & 0.77 & 90.67 & 0.84 & 88.39 & 1.18 & 89.34 & 1.19 & 88.78 & 1.33 & 90.56 & 0.85 & 89.85 & 1.03 \\
     & \textbf{DCA-Net} & \textbf{91.83} & \textbf{0.72} & \textbf{91.59} & \textbf{0.81} & \textbf{89.93} & \textbf{0.77} & \textbf{91.99} & \textbf{0.64} & \textbf{90.68} & \textbf{0.93} & \textbf{90.57} & \textbf{0.82} & \textbf{90.93} &
     \textbf{0.78}\\
     \hline
     \multirow{9}{0.4in}{Unseen} & DeepAll & 87.87 & 2.05 & 85.37 & 1.82 & 82.49 & 2.97 & 86.87 & 2.25 & 84.48 & 2.18 & 85.58 & 1.82 & 85.52 & 2.18 \\
     & BigAug & 88.62 & 1.70 & 86.22 & 1.56 & 83.76 & 2.72 & 87.35 & 1.98 & 85.53 & 1.90 & 85.83 & 1.75 & 86.21 & 1.93 \\
     & MASF & 88.70 & 1.69 & 86.20 & 1.54 & 84.16 & 2.39 & 87.43 & 1.91 & 86.18 & 1.85 & 86.57 & 1.47 & 86.55 & 1.81 \\ 
     & SAML & 88.89 & 1.38 & 87.17 & 1.46 & \textbf{85.60} & 2.07 & 86.96 & 1.56 & 86.19 & 1.77 & 88.12 & 1.22 & 87.16 & 1.58 \\
     \cline{2-16}
     & DCA-Net$^*$ & 90.24 & 1.15 & 88.12 & 1.16 & 82.45 & 1.76 & 88.28 & 1.23 & 86.14 & 1.64 & 88.90 & 0.97 & 87.36 & 1.32 \\
     & DCA-Net(N=4) & 90.22 & 1.19 & 87.06 & 1.31 & 83.62 & 1.61 & 88.03 & 1.29 & 86.03 & 1.84 & \textbf{89.97} & \textbf{0.89} & 87.49 & 1.36 \\
	 & DCA-Net(N=16) & 90.24 & \textbf{1.11} & 87.24 & 1.31 & 84.18 & 1.90 & 88.64 & 1.18 & 86.46 & 1.64 & 88.76 & 1.02 & 87.59 & 1.36 \\
	 & \textbf{DCA-Net(ours)} & \textbf{90.61} & 1.12 & \textbf{88.31} & \textbf{1.14} & 84.89 & \textbf{1.76} & \textbf{89.22} & \textbf{1.09} & \textbf{86.78} & \textbf{1.58} & 89.17 & 1.02 & \textbf{88.16} & \textbf{1.29} \\
     \hline
    \end{tabular}}
\end{table}
\subsubsection{Ablation Study.}To validate the role of our $L_{div}$, we removed this term in the loss function and used $\mathcal{L}_{seg}$ with $\mathcal{L}_{comp}$ to train DCA-Net, and the corresponding model is referred to as  DCA-Net$^{*}$ . The results in Table~\ref{tab2} show that DCA-Net$^{*}$ already outperformed DeepAll, BigAug, MASF, and SAML, which achieves 0.2\% average Dice and 0.26 $mm$ ASD improvements compared with that of SAML, with average Dice and ASD of 87.36\% and 1.32 $mm$, respectively. Introducing $L_{div}$ to DCA-Net further improved the average Dice to 88.16\%, and decreased the average ASD to 1.29~$mm$, demonstrating the importance of learning divergent basis representations for our PDP module. Besides, we investigated the effect of the number of basis preceptors of PDP (i.e., size of domain representation bank)  by setting $N$ to 4, 8 and 16 respectively. Results in  Table~\ref{tab2} show that `DCA-Net(ours)' with $N=8$ achieved the best performance. This demonstrates that a small number of basis preceptors is not enough to represent the different domains on multi-site datesets, while a very large number of basis preceptors or domain representation bank size does not lead to much gain in performance.
\subsubsection{Performance on Seen Domains.} We also investigated the performance of our method on the seen source domains. We followed~\cite{liu2020shape} to split each source domain dataset at patient level into 70\%, 10\% and 20\% for training, validation and testing. 
We retrained DeepAll with $\mathit{L}_{seg}$ and $\mathcal{L}_{comp}$ under the new data setting, which is referred to as DeepAll$^*$, and compared with `Intra-site' that means training and testing 6 domain-specific models, where each model only uses the data from a single domain. The `seen' section of Table~\ref{tab2} shows that Intra-site got a good performance by requiring each domain to provide images for training. DeepAll$^*$ can improve the performance due to the access to a larger dataset from multiple sites. What's more, DCA-Net outperformed DeepAll$^*$, which demonstrates that the proposed domain composition and attention method also improves the segmentation performance on multiple seen domains.
\section{Conclusion}
We present a Domain Composition and Attention-based model (DCA-Net) to tackle the commonly faced domain generalization problem in medical image segmentation. we propose a Parallel Domain Preceptor (PDP) that synchronously uses domain preceptors to construct a domain representation bank with as set of basis domain representations. Then, the domain attention module learns to predict coefficients for a linear combination of basis representations from the representation bank, which is used to calibrate and neutralize the domain bias of the input feature. Meanwhile, we introduce a novel divergence constraint function to guide the preceptors to capture divergent domain representations, which is important for imroving the generalizability. Experimental results showed the effectiveness of our proposed DCA-Net for achieving robust performance in prostate segmentation from multiple sites and even unseen domains. In the future, it is of interest to apply our DCA to other backbone networks and validate it with other segmentation tasks with more seen and unseen domains.

\section{Acknowledgement}
This work was supported by the National Natural Science Foundations of China [61901084 and 81771921] funding, key research and development project of Sichuan province, China [No. 20ZDYF2817].
\bibliographystyle{splncs04}
\bibliography{domain}

\begin{thebibliography}{10}
\providecommand{\url}[1]{\texttt{#1}}
\providecommand{\urlprefix}{URL }
\providecommand{\doi}[1]{https://doi.org/#1}

\bibitem{balaji2018metareg}
Balaji, Y., Sankaranarayanan, S., Chellappa, R.: Metareg: Towards domain
  generalization using meta-regularization. In: Proceedings of the 32nd
  International Conference on Neural Information Processing Systems. pp.
  1006--1016 (2018)

\bibitem{chen2018semantic}
Chen, C., Dou, Q., Chen, H., Heng, P.A.: Semantic-aware generative adversarial
  nets for unsupervised domain adaptation in chest x-ray segmentation. In: Shi,
  Y., Suk, H.I., Liu, M. (eds.) Machine Learning in Medical Imaging. pp.
  143--151. Springer International Publishing, Cham (2018)

\bibitem{chen2016attention}
Chen, L.C., Yang, Y., Wang, J., Xu, W., Yuille, A.L.: Attention to scale:
  Scale-aware semantic image segmentation. In: Proceedings of the IEEE
  conference on computer vision and pattern recognition. pp. 3640--3649 (2016)

\bibitem{dou2019domain}
Dou, Q., Castro, D.C., Kamnitsas, K., Glocker, B.: Domain generalization via
  model-agnostic learning of semantic features. arXiv preprint arXiv:1910.13580
   (2019)

\bibitem{dou2018unsupervised}
Dou, Q., Ouyang, C., Chen, C., Chen, H., Heng, P.A.: Unsupervised
  cross-modality domain adaptation of convnets for biomedical image
  segmentations with adversarial loss. arXiv preprint arXiv:1804.10916  (2018)

\bibitem{ganin2016domain}
Ganin, Y., Ustinova, E., Ajakan, H., Germain, P., Larochelle, H., Laviolette,
  F., Marchand, M., Lempitsky, V.: Domain-adversarial training of neural
  networks. The journal of machine learning research  \textbf{17}(1),
  2096--2030 (2016)

\bibitem{gu2021comprehensive}
{Gu}, R., {Wang}, G., {Song}, T., {Huang}, R., {Aertsen}, M., {Deprest}, J.,
  {Ourselin}, S., {Vercauteren}, T., {Zhang}, S.: Ca-net: Comprehensive
  attention convolutional neural networks for explainable medical image
  segmentation. IEEE Transactions on Medical Imaging  \textbf{40}(2),  699--711
  (2021)

\bibitem{hu2018se}
Hu, J., Shen, L., Sun, G.: Squeeze-and-excitation networks. In: Proceedings of
  the IEEE Conference on Computer Vision and Pattern Recognition (CVPR) (June
  2018)

\bibitem{kamnitsas2017unsupervised}
Kamnitsas, K., Baumgartner, C., Ledig, C., Newcombe, V., Simpson, J., Kane, A.,
  Menon, D., Nori, A., Criminisi, A., Rueckert, D., et~al.: Unsupervised domain
  adaptation in brain lesion segmentation with adversarial networks. In:
  International conference on information processing in medical imaging. pp.
  597--609. Springer (2017)

\bibitem{li2018learning}
Li, D., Yang, Y., Song, Y.Z., Hospedales, T.: Learning to generalize:
  Meta-learning for domain generalization. In: Proceedings of the AAAI
  Conference on Artificial Intelligence. vol.~32 (2018)

\bibitem{li2018domain}
Li, H., Pan, S.J., Wang, S., Kot, A.C.: Domain generalization with adversarial
  feature learning. In: Proceedings of the IEEE Conference on Computer Vision
  and Pattern Recognition. pp. 5400--5409 (2018)

\bibitem{litjens2014evaluation}
Litjens, G., Toth, R., van~de Ven, W., Hoeks, C., Kerkstra, S., van Ginneken,
  B., Vincent, G., Guillard, G., Birbeck, N., Zhang, J., et~al.: Evaluation of
  prostate segmentation algorithms for mri: the promise12 challenge. Medical
  image analysis  \textbf{18}(2),  359--373 (2014)

\bibitem{liu2020shape}
Liu, Q., Dou, Q., Heng, P.A.: Shape-aware meta-learning for generalizing
  prostate mri segmentation to unseen domains. In: International Conference on
  Medical Image Computing and Computer-Assisted Intervention. pp. 475--485.
  Springer (2020)

\bibitem{liu2020ms}
Liu, Q., Dou, Q., Yu, L., Heng, P.A.: Ms-net: Multi-site network for improving
  prostate segmentation with heterogeneous mri data. IEEE transactions on
  medical imaging  \textbf{39}(9),  2713--2724 (2020)

\bibitem{oktay2018attention}
Oktay, O., Schlemper, J., Folgoc, L.L., Lee, M., Heinrich, M., Misawa, K.,
  Mori, K., McDonagh, S., Hammerla, N.Y., Kainz, B., et~al.: Attention u-net:
  Learning where to look for the pancreas. arXiv preprint arXiv:1804.03999
  (2018)

\bibitem{perone2019}
Perone, C.S., Ballester, P., Barros, R.C., Cohen-Adad, J.: Unsupervised domain
  adaptation for medical imaging segmentation with self-ensembling. NeuroImage
  \textbf{194},  1--11 (2019)

\bibitem{ronneberger2015u}
Ronneberger, O., Fischer, P., Brox, T.: U-net: Convolutional networks for
  biomedical image segmentation. In: International Conference on Medical image
  computing and computer-assisted intervention. pp. 234--241. Springer (2015)

\bibitem{shen2017deep}
Shen, D., Wu, G., Suk, H.I.: Deep learning in medical image analysis. Annual
  review of biomedical engineering  \textbf{19},  221--248 (2017)

\bibitem{wang2020dofe}
Wang, S., Yu, L., Li, K., Yang, X., Fu, C.W., Heng, P.A.: Dofe: Domain-oriented
  feature embedding for generalizable fundus image segmentation on unseen
  datasets. IEEE Transactions on Medical Imaging  (2020)

\bibitem{wang2019towards}
Wang, X., Cai, Z., Gao, D., Vasconcelos, N.: Towards universal object detection
  by domain attention. In: Proceedings of the IEEE/CVF Conference on Computer
  Vision and Pattern Recognition. pp. 7289--7298 (2019)

\bibitem{zhang2020generalizing}
Zhang, L., Wang, X., Yang, D., Sanford, T., Harmon, S., Turkbey, B., Wood,
  B.J., Roth, H., Myronenko, A., Xu, D., et~al.: Generalizing deep learning for
  medical image segmentation to unseen domains via deep stacked transformation.
  IEEE transactions on medical imaging  \textbf{39}(7),  2531--2540 (2020)

\bibitem{zhang2012sparse}
Zhang, S., Zhan, Y., Dewan, M., Huang, J., Metaxas, D.N., Zhou, X.S.: Towards
  robust and effective shape modeling: Sparse shape composition. Medical Image
  Analysis  \textbf{16}(1),  265--277 (2012)

\end{thebibliography}
%




\end{document}